\documentstyle[11pt]{article}

\begin{document}

\begin{flushright}
DPNU-01-03
\end{flushright}
\vspace{-0.7cm}
\begin{center}
\Large\bf
New Pattern of Chiral Symmetry Restoration
\footnote{%
Talk given at YITP workshop
``Fundamental Problems and Applications of Quantum Field Theory''
(December 20 -- 22, 2000, Yukawa Institute, Kyoto, Japan).
This talk is based on the works done in collaboration with
Prof. Yamawaki~\cite{HY:letter, HY:matching, HY:VM}.}
\end{center}

\begin{center}
{\Large Masayasu Harada}\\
{\it
Department of Physics, Nagoya University, Nagoya 464-8602, JAPAN
}
\end{center}

\begin{abstract}
In this talk I summarize our recent
works on the chiral symmetry restoration in the large flavor
QCD using the hidden local symmetry (HLS) as an effective
field theory of QCD.
Bare parameters in the HLS are determined by matching the HLS with
QCD at the matching scale through the Wilsoninan matching conditions.
This Wilsonian matching leads
to the vector manifestation of the Wigner realization
of the chiral symmetry in
which the symmetry is restored 
by the massless degenerate pion (and its flavor
partners) and rho meson (and its flavor partners) as the chiral
partner.
\end{abstract}

Chiral symmetry restoration in the large flavor QCD
is an interesting subject to study.
In Ref.~\cite{BZ} 
an IR fixed point was found to exist
at two-loop beta function
for large $N_f$ ($< \frac{11}{2} N_c$)
in the QCD with massless $N_f$ flavors.
Based on this IR fixed point, it was found~\cite{ATW}
through the modified ladder Schwinger-Dyson equation
that chiral symmetry restoration 
in fact takes place for $N_f$ 
such that $N_f^{\rm cr} < N_f < \frac{11}{2} N_c$, where
$N_f^{\rm cr}\simeq 4 N_c$($=12$ for $N_c=3$).
Moreover, the lattice simulation indicates that the chiral restoration 
does occur at $N_f^{\rm cr} \approx 7$~\cite{IKKSY}.

In Ref.~\cite{HY:letter}
we pointed that the chiral restoration takes place for large $N_f$
also in the hidden local symmetry (HLS)~\cite{BKUYY}
{\it by its own dynamics}.
The
inclusion of the quadratic divergences in the renormalization group
equations (RGE's) was essential 
to obtain the phase transition.
This inclusion is also necessary when we match the HLS with the
underlying QCD through 
the Wilsonian matching which
is shown to work
very well for the realistic case of $N_f=3$.~\cite{HY:matching}

In Ref.~\cite{HY:VM} we
applied the Wilsonian matching for the large flavor QCD,
and proposed ``Vector Manifestation (VM)'' of the chiral symmetry
as a novel manifestation of the Wigner realization
in which the vector meson denoted by $\rho$ ($\rho$ meson and its
flavor partner)
becomes massless at the chiral
phase transition point.  Accordingly, the (longitudinal) $\rho$
becomes the chiral partner of the Nambu-Goldstone (NG) boson 
denoted by $\pi$
(pion and its flavor partners).

In this talk I summarize the Wilsonian matching
and the VM.
For details please see Refs.~\cite{HY:letter,HY:matching,HY:VM}.

Let me start from 
the Wilsonian matching of the HLS with QCD
proposed in Ref.~\cite{HY:matching}.
In the HLS 
axialvector and vector current correlators are
well described by the tree contributions with including
${\cal O}(p^4)$ terms~\footnote{%
  It should be noticed that 
  in the HLS we can perform 
  a systematic low-energy expansion~\cite{Tanabashi,HY:matching,HY}.
}
when the momentum is around the matching scale, $Q^2 \sim \Lambda^2$:
\begin{eqnarray}
\Pi_A^{\rm(HLS)}(Q^2) 
=
\frac{F_\pi^2(\Lambda)}{Q^2} - 2 z_2(\Lambda)
\ ,
\quad
\Pi_V^{\rm(HLS)}(Q^2) 
=
\frac{
  F_\sigma^2(\Lambda)
  \left[ 1 - 2 g^2(\Lambda) z_3(\Lambda) \right]
}{
  M_v^2(\Lambda) + Q^2
} 
- 2 z_1(\Lambda)
\ ,
\label{Pi A V HLS}
\end{eqnarray}
where $g(\Lambda)$ is the bare HLS gauge coupling,
$F_\sigma^2(\Lambda) = a(\Lambda) F_\pi^2(\Lambda)$ is the bare decay
constant of the would-be NG boson $\sigma$ 
absorbed into the HLS gauge boson, and 
$M_v^2(\Lambda) \equiv g^2(\Lambda) F_\sigma^2(\Lambda)$ is the bare
HLS gauge boson mass.
$z_1(\Lambda)$, $z_2(\Lambda)$ and $z_3(\Lambda)$ are the (bare)
coefficients of the relevant ${\cal O}(p^4)$ terms.
On the other hand,
the same correlators are evaluated by the OPE 
up till ${\cal O}(1/Q^6)$~\cite{SVZ}:
\begin{eqnarray}
&&
\Pi_A^{\rm(QCD)}(Q^2) = \frac{1}{8\pi^2}
\Biggl[
  - \left( 1 + \frac{\alpha_s}{\pi} \right) \ln \frac{Q^2}{\mu^2}
  + \frac{\pi^2}{3} 
    \frac{
      \left\langle 
        \frac{\alpha_s}{\pi} G_{\mu\nu} G^{\mu\nu}
      \right\rangle
    }{ Q^4 }
  + \frac{\pi^3}{3} \frac{1408}{27}
    \frac{\alpha_s \left\langle \bar{q} q \right\rangle^2}{Q^6}
\Biggr]
\ ,
\nonumber\\
&&
\Pi_V^{\rm(QCD)}(Q^2) = \frac{1}{8\pi^2}
\Biggl[
  - \left( 1 + \frac{\alpha_s}{\pi} \right) \ln \frac{Q^2}{\mu^2}
  + \frac{\pi^2}{3} 
    \frac{
      \left\langle 
        \frac{\alpha_s}{\pi} G_{\mu\nu} G^{\mu\nu}
      \right\rangle
    }{ Q^4 }
  - \frac{\pi^3}{3} \frac{896}{27}
    \frac{\alpha_s \left\langle \bar{q} q \right\rangle^2}{Q^6}
\Biggr]
\ .
\label{Pi A V OPE}
\end{eqnarray}
where $\mu$ is the renormalization scale of QCD.
The current correlators in the HLS
in Eq.~(\ref{Pi A V HLS}) can be 
matched with those in QCD in Eq.~(\ref{Pi A V OPE}) up till first
derivative at the matching scale $\Lambda$:
\begin{equation}
\Pi_{A}^{\rm (HLS)} - \Pi_{V}^{\rm (HLS)}
=
\Pi_{A}^{\rm (QCD)} - \Pi_{V}^{\rm (QCD)}
\ ,
\quad
\frac{d}{dQ^2} \Pi_{A,V}^{\rm (HLS)}
=
\frac{d}{dQ^2} \Pi_{A,V}^{\rm (QCD)}
\ ,
\quad
\mbox{at} \ Q^2 = \Lambda^2 \ .
\label{WM cond}
\end{equation}
Here the difference of the current correlators in the first equation 
is taken to eliminate the explicit dependence on the renormalization
scale $\mu$ of QCD.
What is shown in Ref.~\cite{HY:matching} is that this Wilsonian
matching together with the Wilsonian RGE's~\cite{HY:letter} works very
well for the realistic case of $N_f=3$.

Now, let me briefly summarize the VM proposed in
Ref.~\cite{HY:VM}.
One important result of the Wilsonian matching 
is that
{\it $F_\pi^2(\Lambda)$ is non-zero even at the critical point}
where the chiral symmetry
is restored with $\left\langle \bar{q} q \right\rangle = 0$.
Then how do we know by the bare parameters defined at $\Lambda$
whether or not the chiral symmetry is restored ?
A clue comes from the fact that
$\Pi_A^{\rm (QCD)}$ 
and $\Pi_V^{\rm (QCD)}$ in Eq.~(\ref{Pi A V OPE})
agree with each other for any value of $Q^2$ at the critical point.
Thus, through the Wilsonian matching,
it is reasonable to require that
$\Pi_A^{\rm (HLS)}$ and $\Pi_V^{\rm (HLS)}$ in Eq.~(\ref{Pi A V HLS})
agree with each other for {\it any value of $Q^2$}.
This agreement is satisfied by the following conditions:
\begin{equation}
g(\Lambda) \rightarrow 0 \ , 
\quad a(\Lambda) = F_\sigma^2(\Lambda)/F_\pi^2(\Lambda)
  \rightarrow 1 \ , \quad
z_1(\Lambda) - z_2(\Lambda) \rightarrow 0 \ .
\label{vector conditions}
\end{equation}
Since $(g,a)=(0,1)$ is the fixed point of the 
RGE's~\cite{HY:letter},
$g(m_\rho)=0$ and $a(m_\rho)=1$
at the $\rho$ mass scale $m_\rho$.
This implies that
$\rho$ becomes massless ($m_\rho = g(m_\rho)F_\sigma(m_\rho)=0$)
with the current coupling equal to
that of $\pi$ ($F_\sigma(m_\rho) = F_\pi(0)$).
This is nothing but the VM of the chiral symmetry
which is accompanied by the
degenerate 
massless $\pi$ and (longitudinal) $\rho$.
A salient feature of the VM is that
{\it $m_\rho$ approaches to zero faster than
$F_\pi$}:~\cite{HY:VM}
\begin{equation}
m_\rho^2/F_\pi^2(0) \rightarrow a(m_\rho) g^2(m_\rho) \rightarrow 0
\ .
\end{equation}

\paragraph{Acknowledgment}
\ \par

I would like to thank Professor Koichi Yamawaki for collaboration in
Ref.~\cite{HY:letter, HY:matching, HY:VM} on which this talk is based.
I would be grateful to the organizers for giving me 
an opportunity to present this talk.

\end{document}